\documentclass[english,aps,preprint]{revtex4}
\usepackage[T1]{fontenc}
\usepackage[latin1]{inputenc}
\usepackage{graphicx}
\usepackage{amssymb}

\makeatletter

\usepackage{babel}
\makeatother
\begin{document}

\title{A Kinetic Approach to the Calculation of Surface Tension in a Spherical
Drop}

\author{Vincenzo Molinari, Domiziano Mostacci}

\email{domiziano.mostacci@mail.ing.unibo.it}

\homepage{http://tori.ing.unibo.it/DIENCA/ita/personale/associati/Mostacci/home.html}

\author{Margherita Premuda}

\affiliation{INFM-BO and Laboratorio di Ingegneria Nucleare di Montecuccolino,
Università degli Studi di Bologna via dei Colli 16, I-40136 Bologna
(ITALY)}

\begin{abstract}
In literature, surface tension has been investigated mainly from a
thermodynamic standpoint, more rarely with kinetic methods. In the
present work, surface tension in drops is studied in the framework
of kinetic theory, starting from the Sutherland approximation to Van
Der Waals interaction between molecules. Surface tension is calculated
as a function of drop radius: it is found that it approaches swiftly
an asymptotic value, for radii of several times the intermolecular
distance. This theoretical asymptotic value is compared to experimental
values for a few liquids, and is found in reasonable agreement.
\end{abstract}
\maketitle

\section{introduction}

The effect of curvature on surface tension has been investigated widely
in the 50's, but mainly within the framework of Thermodynamics (see,
e.g., \cite{key-1,key-2}). Some studies have taken the point of view
of kinetic theory \cite{key-3,key-4}. More recent work investigates
surface tension, in plane geometry, from the point of view of kinetic
theory, making use of the Sutherland potential \cite{key-5}; still
much work is being done taking the Thermodynamics standpoint (see
for instance \cite{key-6,key-7,key-8}). In the present work the problem
is studied through kinetic equations derived in the following simplifying
assumptions: 

1) \textbf{Free volume method} \cite{key-1}. This approximation can
also be bettered including the molecular exchange with the surrounding
atmosphere \cite{key-9}. This effect might form the object of subsequent
work. 

2) \textbf{Uncorrelated molecules}. This assumption is dictated by
two considerations: first, the double distribution function (or pair
distribution function, as it often referred to) is not well known
as it can be calculated exactly only for extremely simplified situations.
Albeit it is almost invariably called upon, still the assumptions
made to calculate it are often such as to render its benefits very
limited; secondly, the importance of correlation can be judged from
the ratio $\gamma$ between the potential energy at the average intermolecular
distance and the average kinetic energy of molecules \cite{key-10,key-11}:
\begin{equation}
\gamma=\frac{\Phi\left(r_{0}\right)}{\frac{3}{2}K_{B}T}\label{eq 1}\end{equation}

The intermolecular distance can be estimated, as usual, from the number
density $n$ as $n^{\frac{1}{3}}$. As will be shown in section IV,
where $\gamma$ is calculated for several cases, this ratio is of
order unity, as might have been expected: therefore, correlation between
molecules might play some role. However, it will be neglected here
as a first approximation. 

3) \textbf{Constant density}. The number density will be considered
constant with radius: this approximation might turn out questionable
at the interface, particularly in view of the fact that surface tension
is, after all, an interface effect. This assumption will be made here,
as a first approximation, and then reviewed in commenting the results. 

4) \textbf{Sutherland potential}. The intermolecular Van der Waals
forces are modeled with the Sutherland potential. This assumption
could be bettered using the full Lennard-Jones potential, however
at the cost of greater mathematical difficulties; on the other hand,
it can be argued that the error introduced using the Sutherland potential,
with suitable parameters, is a reasonable price to pay for the mathematical
simplification, which allows solution in closed form. As will be shown
in the last section, the results obtained are in very reasonable agreement
with experimental data. 

In the following sections, a simple method will be derived within
the approximations discussed above, to obtain a simple expression
for the surface tension as a function of radius. As the radius becomes
larger, the value of surface tension approaches that of a plane surface.
Calculations presented for several different liquids, in the large
radius limit, are consistent with experimental results for plane surfaces.

\section{Governing equations}

Consider a spherical droplet of radius R centered in the point O,
and a point P at a distance r from O, with r$\in$ {[}0,R{]}. A system
of spherical coordinates can be defined, with origin in P and the
O-P direction as the polar axis, as depicted in Fig. 1. The $\varphi$=0
half-plane can be chosen arbitrarily, due to the spherical symmetry
of the system. 

\begin{figure}[h]
\includegraphics[%
  scale=0.8]{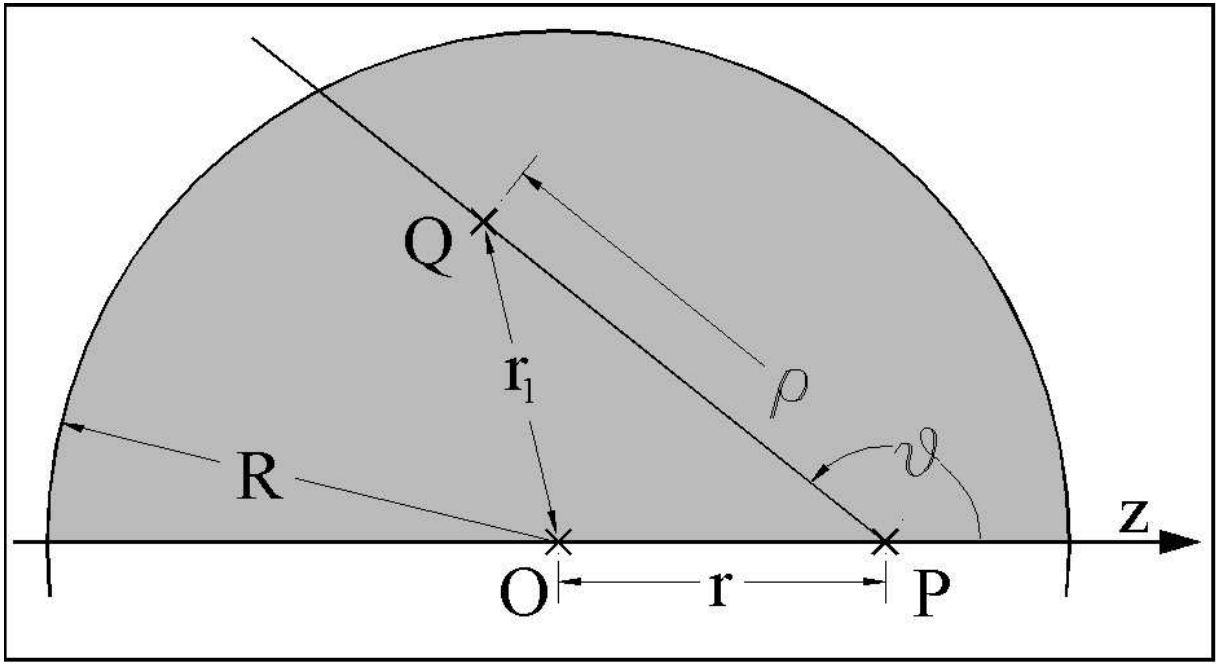}

\caption{Geometry of the problem - section of the droplet on the half-plane
$\varphi$=0}
\end{figure}

Any point Q within the drop that lies on the $\varphi$=0 half-plane
can then be described uniquely with the pair of coordinates $\rho$
and $\vartheta$ (or $\mu=\cos\vartheta$) as defined in Fig. 1.

If forces between particles are purely central forces, i.e., depending
only on a given power $\alpha$ of the distance, then the force due
to the attraction of a particle in Q on the particle in P can be written
as \begin{equation}
\mathbf{F}_{QP}=\frac{G}{\rho^{\alpha}}\hat{\rho}\label{eq:2}\end{equation}
as it is directed from P to Q. In spherical coordinates, the half
circle on the $\varphi$=0 half-plane depicted in Fig. 1 is described
by the equation \begin{equation}
\rho=\sqrt{\left(\mu r\right)^{2}+\left(R^{2}-r^{2}\right)^{2}}-\mu r\label{eq:3}\end{equation}
 or equivalently \begin{equation}
\mu=\frac{\left(R^{2}-r^{2}\right)-\rho^{2}}{2r\rho}\label{eq:4}\end{equation}
 therefore the domain corresponding to the intersection of the drop
with the half-plane is defined as \begin{equation}
\mu\in\left[-1,1\right]\quad;\quad\rho\in\left[0,\sqrt{\left(\mu r\right)^{2}+\left(R^{2}-r^{2}\right)^{2}}-\mu r\right]\label{eq:5}\end{equation}
 or equivalently \begin{equation}
\rho\in\left[0,R+r\right]\quad;\quad\mu\in\left[-1,min\left\{ 1,\,\frac{\left(R^{2}-r^{2}\right)-\rho^{2}}{2r\rho}\right\} \right]\label{eq:6}\end{equation}
 If there are $n\left(r_{1}\right)$ particles per unit volume (due
to the spherical symmetry the density can only depend on the radial
position $r_{1}$, and this can be expressed in terms of r, $\rho$
and $\mu$), in a volume element dV (expressed in spherical coordinates)
at location $\left(\rho,\mu,\varphi\right)$ there are a number of
particles given by\begin{equation}
ndV=n\left(r_{1}\left[r,\rho,\mu\right]\right)\rho^{2}d\rho d\mu d\varphi\label{eq:7}\end{equation}
 exerting on the particle in P a force $d\mathit{\mathbb{\mathcal{F}}}$
\begin{equation}
d\mathcal{F}=\frac{G}{\rho^{\alpha}}ndV=Gn\left(r_{1}\right)\rho^{2-\alpha}d\rho d\mu d\varphi\label{eq:8}\end{equation}
 The z-component of $d\mathit{\mathbb{\mathcal{F}}}$ is \begin{equation}
d\mathit{\mathbb{\mathcal{F}}}_{z}=d\mathit{\mathbb{\mathcal{F}}}\cos\vartheta=Gn\left(r_{1}\right)\rho^{2-\alpha}\mu d\rho d\mu d\varphi\label{eq:9}\end{equation}
 and in the same vein the component perpendicular to z: \begin{equation}
d\mathit{\mathbb{\mathcal{F}}}_{\perp}=d\mathit{\mathbb{\mathcal{F}}}\sin\vartheta=Gn\left(r_{1}\right)\rho^{2-\alpha}\sqrt{1-\mu^{2}}d\rho d\mu d\varphi\label{eq:10}\end{equation}
 The first can be integrated readily with respect to $\varphi$, yielding
\begin{equation}
\int_{0}^{2\pi}d\mathit{\mathbb{\mathcal{F}}}_{z}=2\pi Gn\left(r_{1}\right)\rho^{2-\alpha}\mu d\rho d\mu\label{eq:11}\end{equation}
 As for the second, some further discussion is appropriate. To evaluate
the surface tension one can proceed as follows: imagine cutting the
drop in two halves, each one exerting a certain amount of attraction
on the other. Consider now a molecule that is lying right on the edge
of one half drop, say at the intersection of the polar axis with the
drop surface. The component perpendicular to the axis, and lying on
on the $\varphi=0$ plane, of the force acting on this molecule that
is due to an element of volume at location ($\rho$,$\vartheta$,$\varphi$)
is given by \begin{equation}
d\mathit{\mathbb{\mathcal{F}}}_{\perp}\cos\varphi=Gn\left(r_{1}\right)\rho^{2-\alpha}\sqrt{1-\mu^{2}}d\rho d\mu\cos\varphi d\varphi\label{eq:12}\end{equation}
The tangential force $dF_{T}$ acting on this molecule due to all
elements with coordinates ($\rho$,$\vartheta$), is given by the
integral over $\varphi$ of $d\mathit{\mathbb{\mathcal{F}}}_{\perp}\cos\varphi$
in one of the halves of the drop, e.g., for $\varphi\in\left[-\frac{\pi}{2},\frac{\pi}{2}\right]$:
\begin{equation}
dF_{T}=\int_{-\frac{\pi}{2}}^{+\frac{\pi}{2}}\left[Gn\left(r_{1}\right)\rho^{2-\alpha}\sqrt{1-\mu^{2}}d\rho d\mu\right]\cos\varphi d\varphi=2Gn\left(r_{1}\right)\rho^{2-\alpha}\sqrt{1-\mu^{2}}d\rho d\mu\label{eq:13}\end{equation}
 To calculate the total tangential force on the molecule, the above
expression is to be integrated over $\rho$ and $\mu$. In the following
the density will be considered constant, as discussed in point 3)
in the introduction. Recalling the expression for the domain in Eq.
(\ref{eq:6}), noted that in this case $r=R$, the following integral
is obtained: \begin{equation}
\frac{F_{T}}{2nG}=\int_{D}^{2R}\rho^{2-\alpha}d\rho\int_{-1}^{\frac{-\rho}{2R}}\sqrt{1-\mu^{2}}d\mu=\frac{1}{2}\int_{D}^{2R}\rho^{2-\alpha}\left\{ \arcsin\left(\frac{-\rho}{2R}\right)-\frac{\rho}{2R}\sqrt{1-\left(\frac{\rho}{2R}\right)^{2}}+\frac{\pi}{2}\right\} d\rho\label{eq:14}\end{equation}
Here D is the distance of minimum approach, as will be discussed in
the next section.

\section{Solution with Van der Waals type forces}

Van der Waals interactions are usually represented as a Lennard-Jones
potential \cite{key-12} \begin{equation}
U_{LJ}\left(r\right)=4\varepsilon\left[\left(\frac{\sigma}{r}\right)^{12}-\left(\frac{\sigma}{r}\right)^{6}\right]\label{eq:15}\end{equation}
 This last can be approximated by a Sutherland potential \cite{key-12},
usually written as \begin{equation}
U_{S}\left(r\right)=\left\{ \begin{array}{ccc}
\infty &  & r<D\\
-A\left(\frac{D}{r}\right)^{\omega} &  & r\geq D\end{array}\right.\label{eq:16}\end{equation}
 with suitable choice of the (positive) constants $\omega$, A and
D. A condition typically imposed is to conserve the asymptotic behavior
of ULJ, which entails $\omega=6$ and $A\cdot D^{6}=4\varepsilon\cdot\sigma^{6}$;
a second condition can be chosen as follows: the rest position of
a particle in the field of another particle is the bottom of the potential
well, therefore it seems reasonable to choose the constants so that
this position is the same in the two cases. Therefore, $D=\sigma\cdot2^{\frac{1}{6}}$.
With the above positions the Sutherland potential becomes \begin{equation}
U_{S}\left(r\right)=\left\{ \begin{array}{ccc}
\infty &  & r<D=\sigma\cdot2^{\frac{1}{6}}\\
-\varepsilon\left(\frac{D}{r}\right)^{\omega} &  & r\geq D=\sigma\cdot2^{\frac{1}{6}}\end{array}\right.\label{eq:17}\end{equation}
 This potential is depicted in Fig. 2. This corresponds to the present
model for the force Eq. (\ref{eq:2}), with $\varphi$=7 and\begin{equation}
G=6\varepsilon D^{6}\label{eq:18}\end{equation}

\begin{figure}[h]
\includegraphics[%
  scale=1.5]{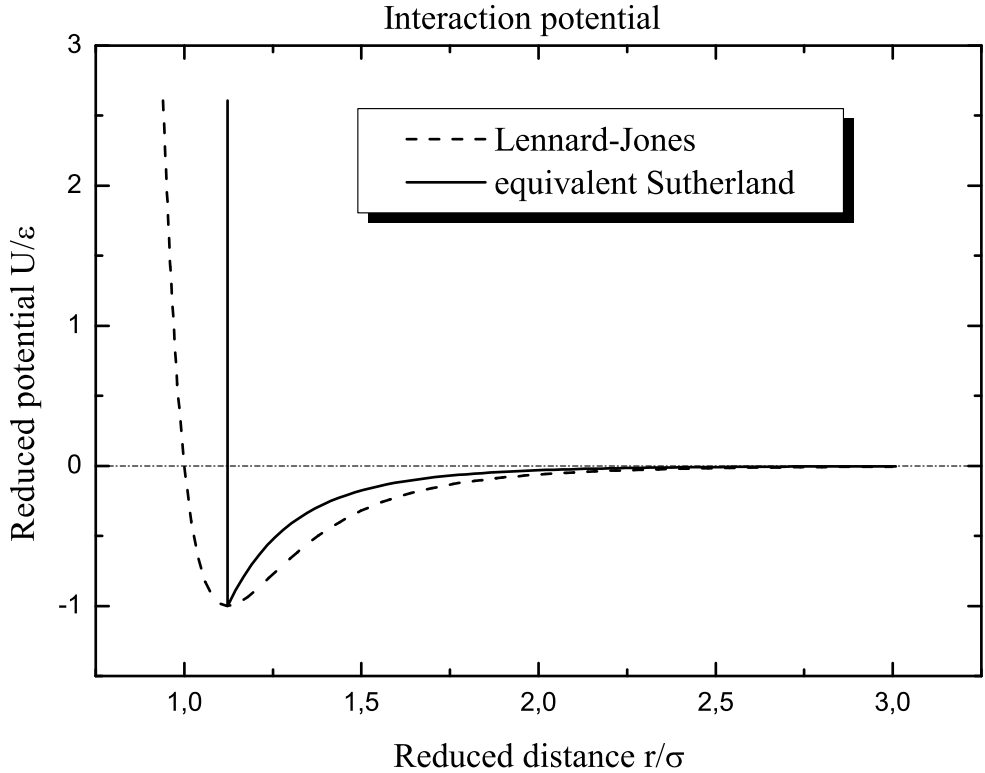}

\caption{Schematized Sutherland potential}
\end{figure}

In the Sutherland approximation, D is the distance of minimum approach,
the lower limit of the integral in Eq. (\ref{eq:14}). Calculating
that integral, and recalling the expression for G in Eq. (\ref{eq:18}),
the following expression is obtained for the tangential force acting
on a molecule: \begin{equation}
F_{T}=\frac{3}{2}n\varepsilon D^{2}\left\{ \arcsin\left(\frac{-D}{2R}\right)+\frac{D}{2R}\sqrt{1-\left(\frac{D}{2R}\right)^{2}}\left[1-2\left(\frac{D}{2R}\right)^{2}\right]+\frac{\pi}{2}\right\} \label{eq:19}\end{equation}
The number of molecule per unit length can be estimated as $n^{\frac{1}{3}}$.
Therefore the force per unit length can be calculated as \begin{equation}
T_{sur}\left(R\right)=\frac{3}{2}n^{\frac{4}{3}}\varepsilon D^{2}\left\{ \arcsin\left(\frac{-D}{2R}\right)+\frac{D}{2R}\sqrt{1-\left(\frac{D}{2R}\right)^{2}}\left[1-2\left(\frac{D}{2R}\right)^{2}\right]+\frac{\pi}{2}\right\} \label{eq:20}\end{equation}
 The graph of the reduced surface tension $\frac{T_{sur}}{\frac{3}{2}n^{\frac{4}{3}}\varepsilon D^{2}}$
as a function of $\frac{2R}{D}$, the drop diameter in units of D,
is shown in Fig. 3

\begin{figure}[h]
\includegraphics[%
  scale=1.5]{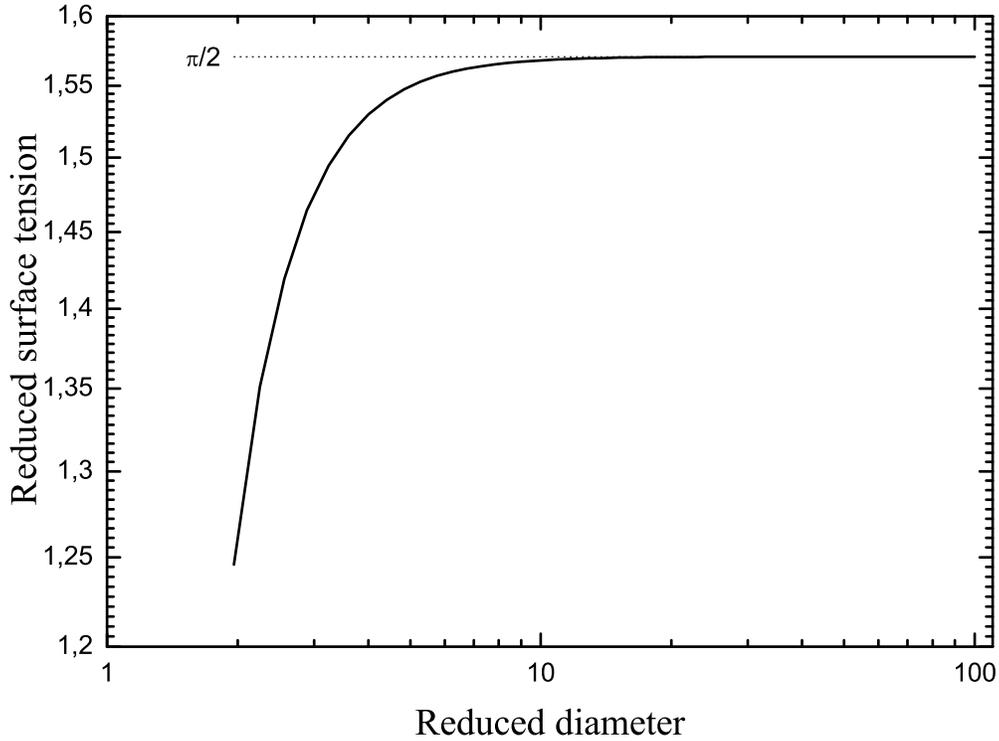}

\caption{Reduced surface tension vs. drop diameter in units of D}
\end{figure}

As can be gathered from the graph, the surface tension approaches
rapidly the limiting value of as the drop diameter grows to only a
few tens of times D.

\section{Comparison with experimental data}

Limiting values of surface tension for large radii have been calculated
from Eq. (\ref{eq:20}) for several liquids, and compared with known
experimental values for plane surfaces. The results are shown in Table
1. As can be gathered from the data reported, the agreement is quite
good for such a qualitative approach, being always within a factor
of 2 to 3 (3.4 for water). As the results depend on the parameters
of the Sutherland potential, agreement might be improved by a different
choice, the one choice made here being somewhat arbitrary albeit reasonable.
In \cite{key-15}, for instance, detailed calculation and comparison
with Lennard-Jones parameters is reported for several inert gases.

\bigskip{}
\includegraphics[%
  scale=0.78]{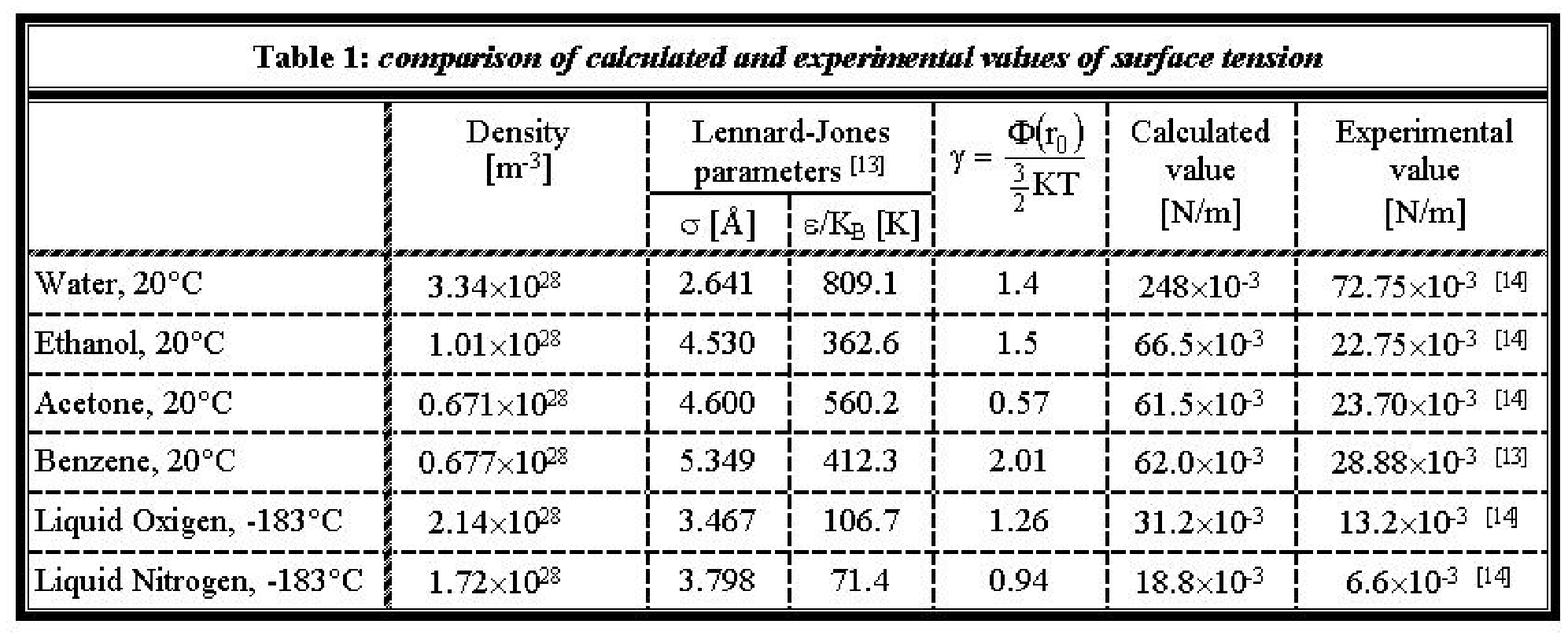}
\smallskip{}

On the other hand, the density is indeed essentially constant throughout
the drop except near the boundary, where it is generally found to
decrease \cite{key-1,key-5,key-16,key-17,key-18,key-19,key-20}. As
the layers close to the boundary give the main contribution to the
integral in Eq. (14), the actual density profile should be considered
in the integration. This will form the object of a subsequent work,
for the present the following view will be taken: in the literature,
the density is always shown to decrease rather abruptly near the interface,
taking a relative value at the interface of roughly one half the bulk
value. Therefore, introducing the actual density in the integral will
decrease this latter by a factor that is between 0.5 and 1. On the
other hand, in calculating the number of molecules per unit length
at the surface, the value of density at the surface must be considered;
altogether then, adding the two effects, the result changes by a factor
of between 0.4 and 0.8, which is the right magnitude to offer the
correction needed to reconcile calculated values with those found
experimentally.
\smallskip{}

\section{Conclusions}

As anticipated in the introduction, aim of this work was to propose
a simple equation for calculating surface tension in drops as a function
of radius. The result is in very reasonable agreement with the experiment.
This result was obtained in the simplifying assumptions discussed
in the introduction. Accurate description of the density profile is
needed to improve confidence in the results yielded by the method
proposed.

\end{document}